\documentclass[]{aa}
\usepackage{graphicx}
\usepackage{txfonts}

\begin{document}

\title{Non-thermal emission from secondary pairs in close TeV binary systems}

\author{V. Bosch-Ramon\inst{1} \and D. Khangulyan\inst{1} \and F.~A. Aharonian\inst{2,1}} 

\institute{Max Planck Institut f\"ur Kernphysik, Saupfercheckweg 1, Heidelberg 69117, Germany; 
vbosch@mpi-hd.mpg.de; Dmitry.Khangulyan@mpi-hd.mpg.de; Felix.Aharonian@dias.ie \and
Dublin Institute for Advanced Studies, Dublin, Ireland}

\offprints{ \\ \email{vbosch@mpi-hd.mpg.de}}

\abstract
{Massive hot stars produce dense ultraviolet (UV) photon fields in their surroundings. If a very high-energy (VHE) gamma-ray 
emitter is located close
to the star, then gamma-rays are absorbed in the stellar photon 
field, creating secondary (electron-positron) pairs.}{We study the broadband emission of these secondary pairs in the stellar
photon and magnetic fields.}{Under certain assumptions on the stellar wind and the magnetic field in the surroundings of a 
massive hot star, we calculate the steady state energy distribution of secondary pairs created 
in the system and its radiation from radio to gamma-rays.}{Under the ambient magnetic field, possibly high enough to
suppress electromagnetic (EM) cascading, the energy of secondary pairs is
radiated via synchrotron and single IC scattering producing radio-to-gamma-ray radiation. 
The synchrotron spectral energy distribution (SED) is hard, peaks around X-ray energies, and becomes softer. 
The IC SED is hard as well and peaks around $10$~GeV, becoming also softer at higher energies due to synchrotron
loss dominance.}{The radio emission from secondary pairs is moderate and detectable as a point-like and/or extended source. In
X-rays, the secondary pair synchrotron component may be dominant. At energies $\la 10$~GeV, the secondary pair IC radiation may
be dominant over the primary gamma-ray emission, 
and possibly detectable by the next generation of instruments.}

\maketitle

\section{Introduction}

Recently, several compact TeV emitters harboring an OB type star have been found in our Galaxy: 
PSR~B1259$-$63 is a massive binary system containing a young non-accreting pulsar detected by HESS
(Aharonian et al. \cite{aharonian05a}); LS~5039 is likely a high-mass microquasar also detected by HESS
(Aharonian et~al. \cite{aharonian05b}); LS~I~+61~303 is a high-mass X-ray binary  detected first by MAGIC
(Albert et~al. \cite{albert06}) and recently also by VERITAS (Maier \cite{maier07});  \mbox{Cygnus~X-1} is a
high-mass microquasar harboring a black-hole detected by MAGIC (Albert et~al. \cite{albert07}). In  this
type of objects, a significant fraction of the energy radiated above 100~GeV can be absorbed
via photon-photon interactions  (e.g. Ford \cite{ford84}; Protheroe \& Stanev \cite{protheroe87}; 
Moskalenko \& Karakula \cite{moskalenko94}, Bednarek \cite{bednarek97}; Boettcher \& Dermer
\cite{boettcher05}; Dubus \cite{dubus06b}; Khangulyan et~al.
\cite{khangulyan07}; Reynoso et al. \cite{reynoso08})  producing secondary
pairs that, under the strong radiation and magnetic fields present close to the massive star,  can
radiate efficiently via IC or synchrotron processes. In the case in which IC is the dominant cooling
channel, EM cascades can develop (see, e.g., Aharonian et~al. \cite{aharonian06b}; Bednarek
\cite{bednarek06}; Khangulyan et~al. \cite{khangulyan07}; Orellana et~al. \cite{orellana07}). If
synchrotron emission is the dominant cooling channel, the absorbed radiation is reemitted at lower
energies (e.g. Khangulyan et~al. \cite{khangulyan07}).  We note that the scenario described here could
take place not only in microquasars or non-accreting pulsar systems but in any source harboring a
gamma-ray emitter embedded in a dense photon field.

In this work, we study the radiation of the secondary pairs created by photon-photon absorption of gamma-rays in the surroundings of a 
hot massive star. 
We focus mainly on the
case when the ambient magnetic field is large enough to suppress effectively EM cascading. The structure of the paper goes as follows: in Sect.~\ref{mod}, the adopted model is
described; in Sect.~\ref{res}, the results are presented and discussed, and our conclusions are given in Sect.~\ref{conc}.

\section{A model for pair creation and secondary pair emission}\label{mod}

\subsection{The general picture}

We describe here the physical system formed by an OB star and a gamma-ray emitter separated by a short distance.  
The star is approximated as a point-like source of
(monoenergetic) photons of energy $\epsilon_0$ and luminosity $L_*$. The emitter, 
located at distance $R=d_*$ from the star and assumed to be point-like, 
isotropically produces gamma-rays with an injection
power $L_{\rm \gamma~inj}$, following a power-law distribution of index $\Gamma$. We consider here only the primary radiation above the minimum threshold energy for secondary pair
production, i.e. $\epsilon_{\rm min~th}=1/\epsilon_0$ (in $m_{\rm e} c^2$ units), since we focus on 
the study of the production of secondary pairs and their emission.  The maximum energy of the
primary photons is fixed at 100~TeV. Throughout most of this work, we characterize the star/VHE emitter as follows: $L_*=10^{39}$~erg/s, $d_*=3\times 10^{12}$~cm~s$^{-1}$, and
$\epsilon_0=2\times 10^{-5}$ (corresponding to a star temperature $\approx 40000$~k); we take two values for $\Gamma$, 2 and 3, 
to account either for hard or soft primary 
gamma-ray spectra. 

\begin{figure}[t]
\resizebox{\hsize}{!}{\includegraphics{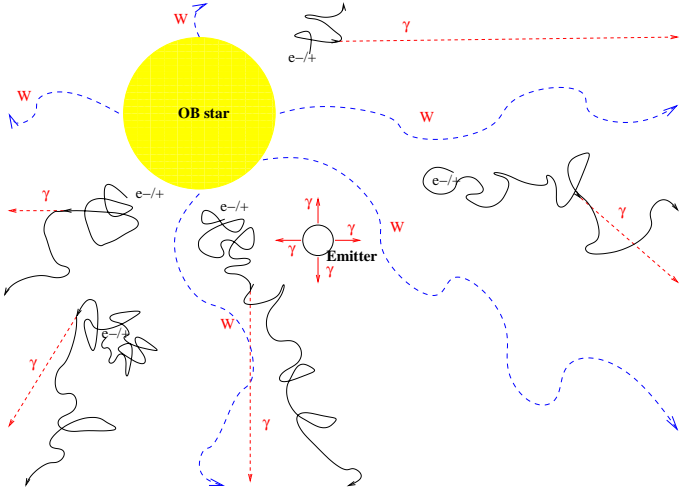}}
\caption{Sketch of the physical scenario considered here. The long-dashed lines represent the wind, the straight dashed lines 
the secondary pair radiation, the straight solid lines the primary gamma-rays, and the curved solid lines the advected/diffusing
secondary pairs.}
\label{picture}
\end{figure}

Important ingredients of the model are the stellar wind and the ambient magnetic field. OB stars present fast winds with
mass loss rates $\dot{M}_{\rm w}\sim 10^{-7}-10^{-5}$~M$_{\odot}$~yr$^{-1}$ and velocities $V_{\rm w}\sim (1-3)\times
10^8$~cm~$s^{-1}$ (e.g. Puls et~al. \cite{puls06}).  We adopt here $\dot{M}_{\rm w}=10^{-6}$~M$_{\odot}$~yr$^{-1}$ and
$V_{\rm w}=2\times 10^8$~cm~s$^{-1}$. Moreover, the star generates a magnetic field in its surroundings with values in the
stellar surface that might be as high as $B_0\sim 1000$~G (e.g. Donati et~al. \cite{donati02}; Hubrig et al.
\cite{hubrig07}).  We assume here that a significant  fraction of the magnetic field is disordered. A sketch of our scenario
is presented in Fig.~\ref{picture}.

\subsection{secondary pairs in the system}

To compute the secondary pair injection spectra in the different regions of the binary system, we use the
anisotropic differential pair production cross section given by eq.~(15) in B\"ottcher \& Schlickeiser
(\cite{boettcher97}). The secondary pair injection spectrum presents a slope similar to that of  the
primary gamma-rays, with the low-energy cutoff similar to the pair creation one, which depends on
the angle between the two incoming photons:
\mbox{$\sim \epsilon_{\rm th}=2/[\epsilon_0(1-\cos\theta_{\gamma\gamma})]$}. The point-like approximation
for the star fails for $R\la 2R_*\sim 20\,R_{\odot}$ 
from the stellar center. For our choice of values of $d_*$, $R_*$,
$L_*$, and $\epsilon_0$, only photons with  energy $>$~few~TeV emitted at an angle $\la 30^\circ$ with
respect to  the star-emitter line can reach the region where the finite size of the star comes into play
(Dubus \cite{dubus06b}). The monoenergetic photon target approximation is roughly valid provided we deal
with a broad distribution of primary gamma-rays. The deviation from the real case is small and will be
neglected here. 

We assume that diffusion takes place in the Bohm regime. Once secondary pairs are injected in different
parts of the system, they isotropize and suffer wind advection, which is the dominant transport mechanism
in the system. Ionization losses in the wind, synchrotron emission in the ambient magnetic field, and IC
scattering of the stellar photons, are the relevant cooling processes of secondary pairs
($\dot{\gamma}$). Relativistic Bremsstrahlung of the secondary pairs in the stellar wind is not
considered since the timescales for this process are much longer than ionization, other radiative, and
advective timescales. All dependencies on $R$ of the relevant physical quantities (the wind density, and
the energy densities of the ambient magnetic and  stellar radiation fields) are taken as $\propto 1/R^2$.
We consider that the TeV emitter produces radiation long enough for the formation of a steady state
distribution of secondary pairs in the system ($n(\gamma,R)$), which can be obtained from the following 
differential equation:  
\begin{equation}  
V_{\rm w}{\partial
n(\gamma,R)\over \partial R}+ {\partial \dot{\gamma}(\gamma,R)n(\gamma,R)\over \partial
\gamma}=q(\gamma,R)\,, \label{eqdif}  
\end{equation} 
where $\gamma$ is the particle Lorentz factor and
$q(\gamma,R)$ is the secondary pair injection rate as a function of energy and distance (derived in the
Appendix).

A restriction on 
$n(\gamma,R)$ is that EM cascades above \mbox{$\sim 100$~GeV} 
must be effectively suppressed by the ambient $B$, which must therefore be above a
certain critical value, $B_{\rm c}$ (see Khangulyan et~al. \cite{khangulyan07}):
\begin{equation}
B_{\rm c}\approx 70\left(L_*\over 10^{39} {\rm  erg/s}\right)^{1/2}\left(R\over R_*\right)^{-1}{\rm G}\ .
\label{eq:b_crit}
\end{equation}
Eq.~(\ref{eq:b_crit}), together with the $R$-dependence assumed here, shows that EM cascades are already
suppressed for a $\approx 20$~G magnetic field at $R=d_*$. This value appears quite moderate when looking
at the possible stellar surface magnetic field values given above. However, even in the case of weaker
magnetic fields, $n(\gamma,R)$ does not
change much if EM cascades are not accounted for $\Gamma\ga 2.5$, since the amount of EM cascade reprocessed energy  
will be relatively small.

For the adopted parameter values in our model, most of the energy of the primary VHE radiation is absorbed for distances
$\la 10^{12}$~cm from the emitter and $\sim d_*$ from the star. Depending on $\Gamma$, radiative cooling leads to an
electron distribution producing synchrotron emission that peaks at either sharply X-ray energies or smoothly in the range X-
to soft gamma-rays. Along with this a fraction of  energy is radiated in the GeV range via IC scattering. The fact that the
minimum energy of the injected secondary pairs is $\sim \epsilon_{\rm min~th}m_{\rm e}c^2$ implies that, even under strong
synchrotron or Thomson IC cooling, the emission of particles with energies below this value will not dominate the total
radiative output. Ionization losses and wind advection lead to steady state secondary pairs radiating very little radio
emission for  $R\la d_*$ from the star. However, the wind transports secondary pair energy to larger $R$, where this energy
can be still efficiently radiated. In these farther regions, synchrotron radio emission can eventually become the dominant
radiative channel. 

\section{Results}\label{res}

\subsection{Injection and evolution of secondary pairs} 

In this section we discuss the spectrum of the secondary pairs injected in the system, $Q_{\rm int}(\gamma)$  $(=\int q(\gamma,R)dR)$, and the
secondary pair energy distribution once the steady state is reached, $N_{\rm int}(\gamma)$ $(=\int n(\gamma,R)dR)$. For illustrative purposes, we perform
our calculations using $L_{\rm \gamma~inj}=3\times 10^{35}$~erg~s$^{-1}$, which is similar to the value
inferred in Khangulyan et~al. (\cite{khangulyan07}) for LS~5039.
We remark that the secondary pair emission luminosity scales linearly with $L_{\rm \gamma~inj}$.

\begin{figure}[t]
\resizebox{\hsize}{!}{\includegraphics{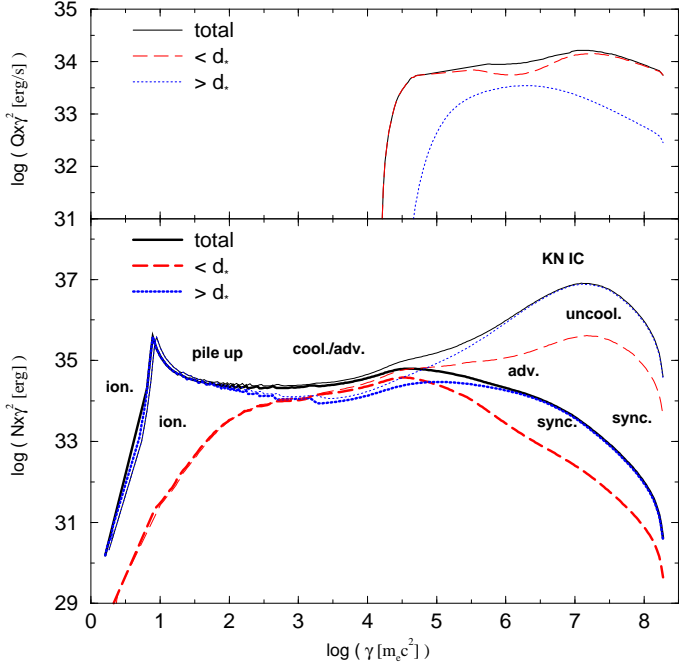}}
\caption{Upper panel: Total $Q_{\rm int}\,\gamma^2$ (solid line). 
$Q_{\rm int}\,\gamma^2$ for $R< d_*$ (long-dashed line) and $> d_*$ (dotted line) are also shown. 
We adopt $\Gamma=2$. Lower panel: 
Total $N_{\rm int}\,\gamma^2$ (solid line). $N_{\rm int}\,\gamma^2$ for
$R< d_*$ (long-dashed line) and $> d_*$ (dotted line) are also shown. We adopt $\Gamma=2$, 
and $B_0=1$ (thin lines) and 100~G 
(thick lines). The dominant cooling mechanisms at different energies are given.}
\label{pairs1}
\end{figure}

\begin{figure}[t]
\resizebox{\hsize}{!}{\includegraphics{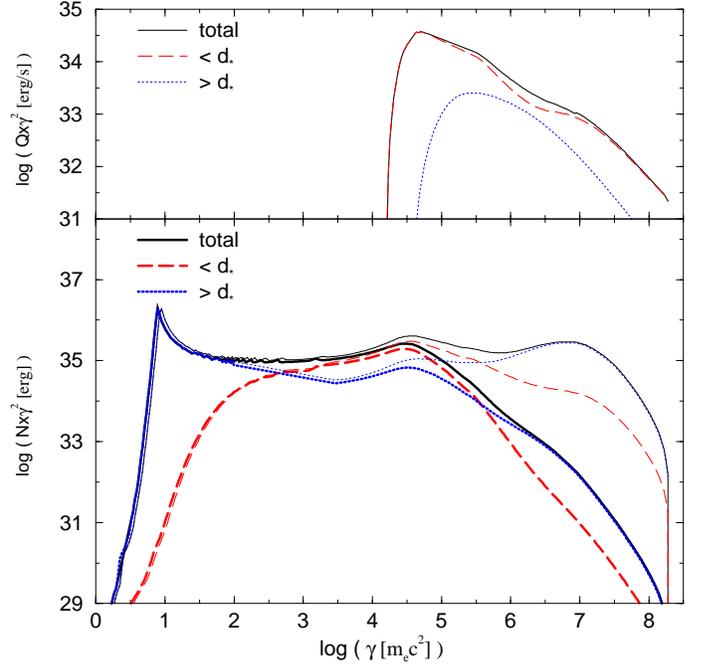}}
\caption{The same as in Fig.~\ref{pairs1} but for $\Gamma=3$.}
\label{pairs2}
\end{figure}

In Figs.~\ref{pairs1} and \ref{pairs2}, upper panels, we show $Q_{\rm int}\,\gamma^2$ for the whole volume, adopting
$\Gamma=2$ and 3, respectively. In the same figures, lower panels, $N_{\rm int}\,\gamma^2$ for the whole region is also
shown. In all the plots, the contributions to the total  $Q_{\rm int}\,\gamma^2$ and  $N_{\rm int}\,\gamma^2$ from $R< d_*$
and $>d_*$ are presented as well. 

As seen in the figures, $Q_{\rm int}(\gamma)$ has a similar shape to that of the primary gamma-rays $\propto
\epsilon_{\gamma}^{-\Gamma}$. As mentioned in Sect.~\ref{mod}, the minimum energy of the injected secondary pairs is the
pair creation threshold energy, i.e.  $\epsilon_{\rm th}m_{\rm e}c^2$, which changes with the angle between the two incoming
photons. As a consequence, both $q(\gamma,R)$ and $n(\gamma,R)$  depend strongly on $R$ as well as on $\Gamma$. In addition,
the shape of $N_{\rm int}(\gamma)$ is related to the dominance of different cooling mechanisms and the impact of wind
advection. 

For particle energies $> \epsilon_{\rm min~th}m_{\rm e}c^2$, depending on $\gamma$, $N_{\rm int}(\gamma)$ is: cooled by
synchrotron losses ($\propto \gamma^{-\Gamma-1}$; for high $B$ in both $R< d_*$ and $> d_*$); uncooled (injection spectrum
slope; for low $B$ and $R< d_*$; because of the fast advection of the particles from this relatively small region); 
or cooled by KN IC ($\propto \gamma^{-\Gamma+1}$; mainly for low $B$ and $R> d_*$;
because there particles have time to lose energy). For particle energies $< \epsilon_{\rm min~th}m_{\rm e}c^2$,
$N_{\rm int}(\gamma)$ is: cooled by synchrotron/Thomson IC ($\propto \gamma^{-2}$; $R< d_*$); advection dominated ($\propto
\gamma^{-2}$; $R> d_*$; because of advection of cooled 
particles created at $R<d_*$); or cooled by ionization
losses ($\propto$~constant; in both $R< d_*$ and $> d_*$). In the range $\gamma\sim 10-100$, the  $1/R^2$-dependence of
losses leads to a strong pileup of particles at $R\gg d_*$, giving a spike feature  in the particle energy distribution. The
cooling mechanisms in different $\gamma$ ranges for $\Gamma=2$ are pointed to in Fig.~\ref{pairs1}, lower panel; for
$\Gamma=3$, despite the softer slope,the pattern is the same.

\begin{figure}[t]
\resizebox{\hsize}{!}{\includegraphics{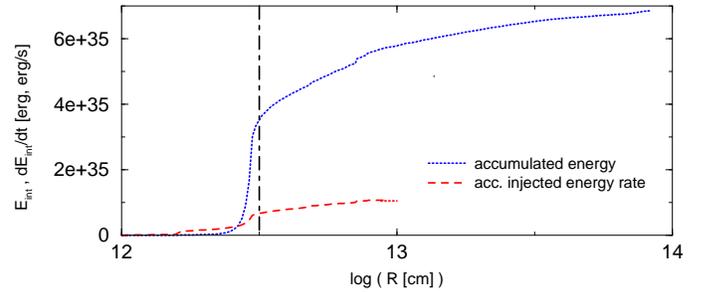}}
\caption{The $R$-distribution of accumulated energy in secondary pairs (long-dashed line), and
the $R$-distribution of accumulated 
energy rate via pair creation (dotted line), are shown. 
The
location of the emitter is marked by a vertical thick dot-dashed line, 
at $\log(d_*)$. In the plot, the
star location would be at the left.}
\label{pairdistr}
\end{figure}

In Fig.~\ref{pairdistr}, we present the total energy ($E_{\rm int}$ in Fig.~\ref{pairdistr}) accumulated
in secondary pairs up to distance $R$, and the total energy rate ($dE_{\rm int}/dt$ in
Fig.~\ref{pairdistr}) from pair creation accumulated up to $R$. The $dE_{\rm int}/dt$ curve is shown up
to the distance at which injection becomes negligeable, i.e. $R_{\rm inj}\sim $~AU. As seen in 
Fig.~\ref{pairdistr}, a significant amount of energy, $\sim 10^{35}$~erg, is transported up to several
AU. Despite this energy budget, the efficiency of the emission from large $R$ will be limited by the low
magnetic and radiation energy densities and the ionization loss dominance. 

\subsection{Emission from secondary pairs}

\begin{figure}[t]
\resizebox{\hsize}{!}{\includegraphics{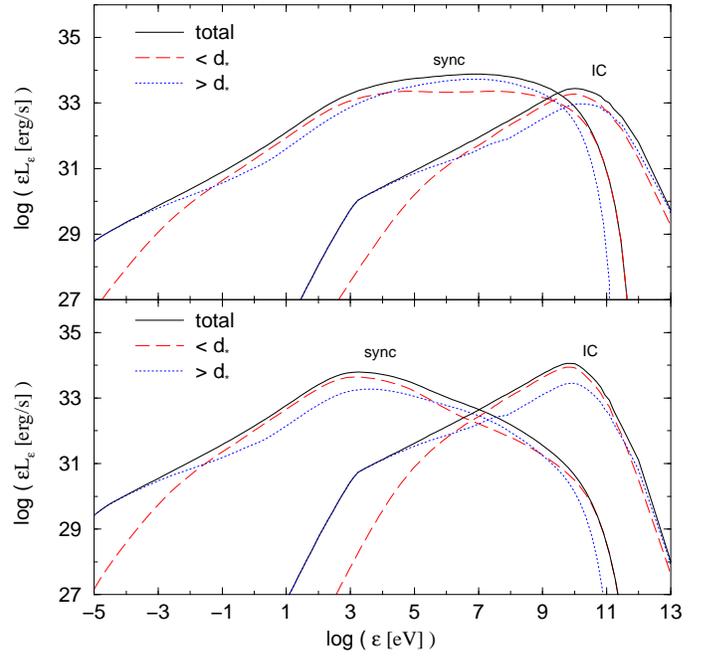}}
\caption{Upper panel: Computed SEDs for the synchrotron and the IC secondary pair emission. 
We adopt $B_0=100$~G and $\Gamma=2$. The
emission from the whole volume (solid line) filled by particles, and that of a 
region located at $R<d_*$ -$\,\sim 0.1$~mas at 3~kpc- (long-dashed line) and 
$>d_*$ (dotted line).
Lower panel: The same as upper-panel but for $\Gamma=3$.}
\label{sed1}
\end{figure}

\begin{figure}[t]
\resizebox{\hsize}{!}{\includegraphics{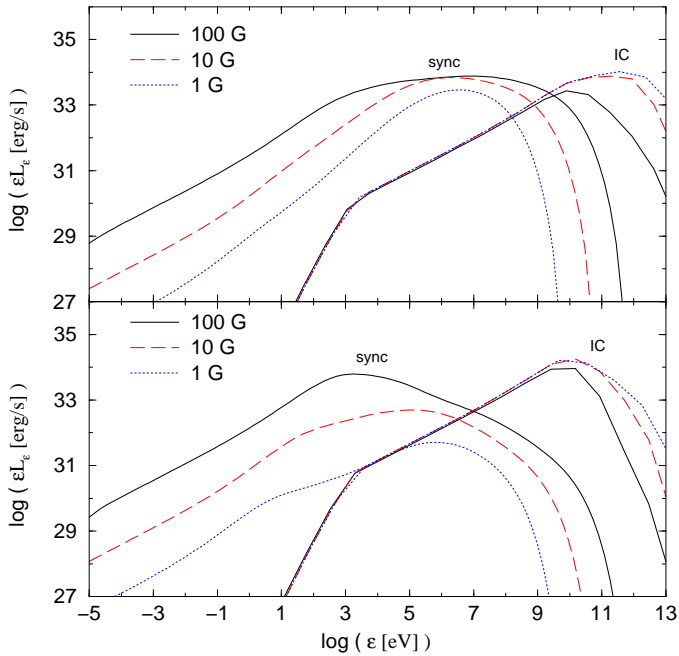}}
\caption{Upper panel: Computed SEDs for the synchrotron and the IC secondary pair emission. 
We adopt $B_0=1$ (dotted line), 
10 (long-dashed line) and 100~G (solid line), 
and $\Gamma=2$.
Lower panel: The same as upper panel but for $\Gamma=3$.}
\label{sed2}
\end{figure}

The calculated SEDs of the synchrotron and IC emission produced by the injected secondary pairs are shown in Figs.~\ref{sed1} and
\ref{sed2}. In Fig.~\ref{sed1}, we have adopted a $B=100$~G, and $\Gamma=2$ (upper panel) and 3 (lower panel). The SEDs of the emission
generated at $R< d_*$ and $> d_*$ from the star are also shown. In Fig.~\ref{sed2}, the  SEDs have been computed adopting $B=1$  (dotted
line), 10 (long-dashed line) and 100~G (solid line), and $\Gamma=2$ (upper panel) and 3 (lower panel). 

For energies of photons produced by secondary pairs with energies  $> \epsilon_{\rm min~th}m_{\rm e}c^2$, the spectrum  becomes softer
for larger $B$ due to the dominating synchrotron channel. For $B=1$~G and $\Gamma=2$, there is a clear hardening in the synchrotron and
IC spectra produced by KN IC cooling (although a proper treatment requires EM cascading). In this photon energy range, the larger $\Gamma$
is, the softer the synchrotron and IC spectra are. 

For energies of photons produced by secondary pairs  with energies $< \epsilon_{\rm min~th}m_{\rm e}c^2$,  the
emission is produced either by synchrotron/Thomson IC cooled secondary pairs ($R\la d_*$), or by secondary
pairs affected already by wind advection ($R>d_*$). The spectral hardening 
at low energies in the synchrotron and IC emission 
for $R\la d_*$ is produced by ionization losses. The particle pileup around
$\gamma\sim 10$ impacts on the low energy radiation from $R> d_*$. 

The inhomogeneity of $q(\gamma,R)$, and the sensitivity of $n(\gamma,R)$ to the cooling and transport
conditions, imply that the produced emission changes strongly with location in the system. Further
complexities like a magnetic field with angular dependences, or a strongly inhomogeneous 
diffusion coefficient,
would yield globally and locally different SEDs. 

\subsubsection{LS~5039 and \mbox{Cygnus~X-1}}

Adopting $B_0\sim 100$~G, our model predicts radio fluxes and X-ray luminosities of several mJy and several
10$^{33}$~erg~s$^{-1}$, respectively. The parameter values adopted in this work are similar to those of
LS~5039 and \mbox{Cygnus~X-1}, making worthy a discussion of the results in the context of these two sources.

The obtained radio fluxes and X-ray luminosities are not far from those found in LS~5039 (e.g. Paredes et~al.
\cite{paredes00}; Bosch-Ramon et~al. \cite{bosch07}), implying that the secondary pair contribution in this
source may be comparable, if not dominant, to that of any intrinsic radio and X-ray component linked to the
TeV emitter itself. We remark that, especially in the radio band, a slightly higher $B$-field would yield
significantly larger radio fluxes due to the $B^2$-dependence of the synchrotron emission. The moderate X-ray
luminosity in LS~5039 permits a look for such a secondary pair component. In the GeV regime, the secondary
pair IC fluxes are far from those found by EGRET (Paredes et~al. \cite{paredes00}), suggesting that the latter
are due to the combination of both primary and secondary pair IC radiation. 

In the case of Cygnus X-1, adopting the same $L_{\rm \gamma~inj}$ as for LS~5039, the predicted radio flux from secondary pairs is a substantial
part of the total radio flux of the source (e.g. Stirling et~al. \cite{stirling01}). In X-rays, explained as thermal emission from
accretion disk/corona-like regions, the high luminosity of \mbox{Cygnus~X-1} makes it impossible observationally to disentangle a possible
secondary pair component. 

The lack of evidence of accretion in LS~5039 (Bosch-Ramon et~al. \cite{bosch07} and references therein) is an important difference
between this source and \mbox{Cygnus~X-1}, which shows clear X-ray accretion features (Albert et~al. \cite{albert07} and references therein).
Nevertheless, as noted, e.g. by Bogovalov \& Kelner (\cite{bogovalov05}), the fact that in some sources the accretion disk luminosity is
undetectable does not imply lack of accretion. This may explain why LS~5039 does not show accretion signatures in its X-ray spectrum. 

\subsubsection{LS~I~+61~303}

\begin{figure*}[t]
\begin{center}
\includegraphics[width=0.68\textwidth,angle=0]{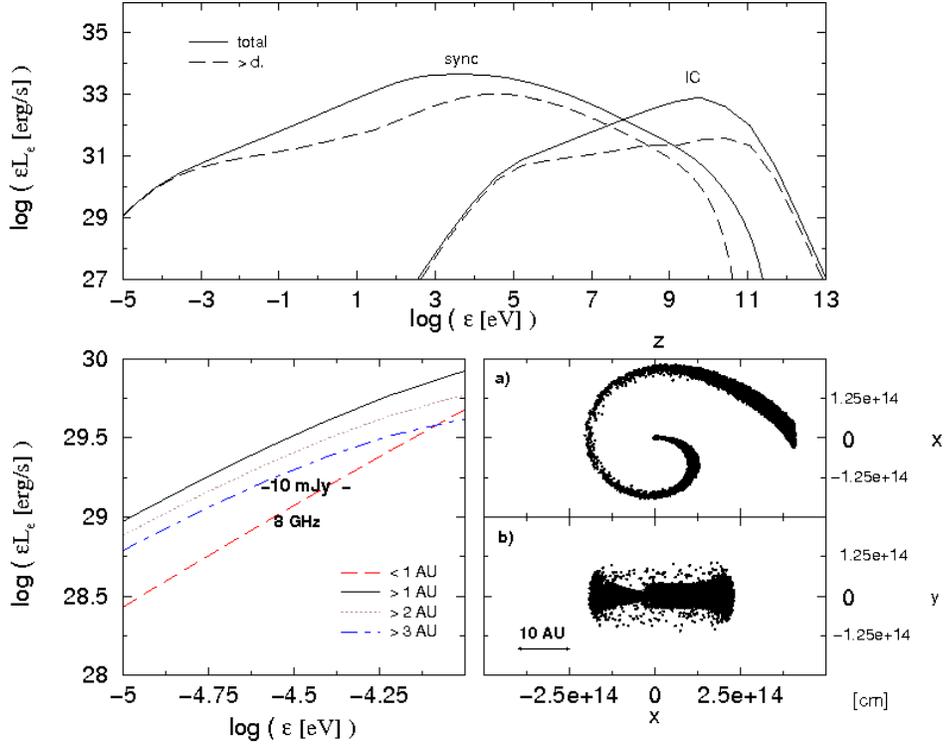}
\caption{Upper panel: Computed SEDs for the synchrotron and the IC secondary pair emission produced in
the whole volume (solid line) and the region beyond $> d_*$ (long dashed). We adopt $B_0=100$~G and $\Gamma=2.5$.
Lower panel, left: Radio SEDs for different regions: $R< 1$ (long dashed), $> 1$ (solid line), $> 2$ (dotted line) and $> 3$~AU 
(dot-dashed line). Lower panel, right: Spatial distribution of the emitting particles, for the orbital plane (a) and a
plane perpendicular to it (b), in the surroundings of LS~I~+61~303 after one orbital period. 
The star is centered in the origin. 
We note that the system size is $\approx 0.4$~AU, and 1~AU$\approx 0.5$~milliarcsecond.}
\label{sed3}
\end{center}
\end{figure*}

For a compact system with the properties listed in Sect.~\ref{mod}, only VLBI interferometers with angular resolution $\la
0.1$~milliarcseconds would be able to resolve the radio emission. Nonetheless, within the constraints of our model, we have
explored  the possibility of explaining the extended radio emission  of a less compact system, LS~I~+61~303, which  presents
a peculiar extended radio structure with changing morphology along the orbit (Dhawan et~al. \cite{dhawan06}).  We have
computed the radio emission produced at the spatial scales of  the observed extended radio structures. Adopting\footnote{The
relevant parameters characterizing the system properties can be found in Bosch-Ramon et~al. (\cite{bosch06}).} $B_0=100$~G,
we have obtained the SEDs for the synchrotron radio emission originated in different regions: $R< 1$, $> 1$, $> 2$, and $>
3$~AU. The radio and broadband (synchrotron plus IC) SEDs are presented in Fig.~\ref{sed3} (lower -left- and upper panels,
respectively); the spatial distribution of the corresponding emitting particles around LS~I~+61~303 after an injection time
of one orbital period is also shown (lower -right- panel).

To implement these calculations, we adopt the orbital distance corresponding to the phases when the source was detected by
MAGIC (Albert et~al. \cite{albert06}): $\sim 6\times 10^{12}$~cm. $\Gamma$ is taken 2.6  (Albert et~al. \cite{albert06}),
and $L_{\rm \gamma~inj}=3\times 10^{35}$~erg~s$^{-1}$, the same as the one taken for LS~5039, enough to explain
observations.  We note that LS~I~+61~303 is not detected by MAGIC above a few hundred GeV when outside the phase range $\sim
0.5-0.7$. Nevertheless, primary gamma-rays with softer spectra may still be injecting a significant amount of secondary
pairs in the system despite being barely detectable above few hundreds GeV. For the spatial distribution of particles,
we followed individually their energy and spatial evolution accounting for advection and (Bohm) diffusion in the wind using
a simple Monte-Carlo simulation (Bosch-Ramon et al., in preparation). The orbital parameters of LS~I~+61~303, like the
eccentricity of the system ($e=0.72$; see Casares et al. \cite{casares05} for the system parameters), were considered.

From Fig.~\ref{sed3}, it is seen that radiation flux levels of $\sim 20$~mJy (8~GHz) are reached.  In this source, most of
the secondary pair radio emission would appear extended, pointing in the direction opposite to the star and bending due to
orbital motion, as shown in the secondary pair spatial distribution in  Fig.~\ref{sed3}. This kind of behavior is similar to
that found by Dhawan et~al. (\cite{dhawan06}) in LS~I~61~+303. These authors associated the radio morphology of this source
to a particular non-accreting pulsar scenario (see Dubus \cite{dubus06a}), although recent hydrodynamical simulations of
stellar/pulsar colliding winds predict quite different morphologies (see Romero et~al. \cite{romero07} and  Bogovalov et~al.
\cite{bogovalov07}).

The predicted X-ray luminosities are similar to those observed in this source (e.g. Sidoli et~al. \cite{sidoli06}), which, as in LS~5039,
is small enough to allow the study of a secondary pair component. Despite the fact that the complexity of the X-ray emission cannot be
completely explained by our model,  the processes considered here  may be behind a significant fraction of the observed X-ray emission.
Like LS~5039, LS~I~+61~303 was also proposed to be a (variable) GeV source (Tavani et~al. \cite{tavani98}), and the secondary pair
contribution to this energy range may be significant (see also Bednarek \cite{bednarek06}).

\section{Conclusions}\label{conc}

We conclude that the presence of a powerful VHE emitter near a massive hot star leads unavoidably 
to non-thermal emission in the stellar wind. The question whether this secondary pair radiation is
detectable depends on the magnetic and radiation field strength in the system. 

In the scenario explored here, we predict moderately hard spectra, and fluxes of several mJy for typical
galactic distances, not far from the fluxes detected in some microquasars. We remark that radio emission
is produced mainly in the regions with  $R> d_*$ and may be dominant over any other radio component (e.g.
linked to the TeV emitter itself), and the radio morphology, flux and spectrum are strongly sensitive to
the geometry and $R$-dependence of the magnetic field. Interestingly, the predicted radio morphology
could resemble that shown by the VHE emitting X-ray binary LS~I~+61~303.

The fluxes of the (likely non-thermal) X-ray emission of several TeV emitting X-ray binaries (e.g.
LS~I~+61~303, LS~5039), typically around 10$^{33}-10^{34}$~erg~s$^{-1}$, are similar to the values
predicted here. It implies that the secondary pair X-ray radiation could be comparable to, or even
dominate over, any intrinsic component linked to the TeV emitter itself. It is worth noting that due to
their moderate X-ray luminosities LS~5039 and LS~I~+61~303 are good candidates to look for a secondary
pair contribution in this energy range. 

We show here that the secondary pair emission at $\sim$~GeV energies, if not strongly dominated by an
intrinsic GeV emitter, could be revealed by GLAST. Above $\epsilon_{\rm min~th}\sim 10$~GeV, soft primary
spectra and/or moderate-to-high magnetic fields would imply low fluxes. In the latter case, significant
synchrotron energy losses would suppress EM cascading. 

Our calculations show that even with simple assumptions on the system geometry, primary gamma-ray
injected spectra, and $B$-field and wind structure, the resulting situation is quite complex. Thus a
detailed characterization of the secondary pair non-thermal emission in a particular source is a
difficult task for which stellar wind physics, primary VHE emission modeling, and high quality data are
required. 

\begin{acknowledgements} 
The authors thank the anonymous referee for constructive comments.
The authors are grateful to Andrew Taylor for a thorough reading of the manuscript.
V.B-R. gratefully acknowledges support from the Alexander von Humboldt Foundation.
V.B-R. acknowledges support by DGI of MEC under grant
AYA2007-68034-C03-01, as well as partial support by the European Regional Development Fund (ERDF/FEDER).
\end{acknowledgements}

{}

\section{Appendix}

\begin{figure}[t]
\begin{center}
\includegraphics[width=0.34\textwidth,angle=0]{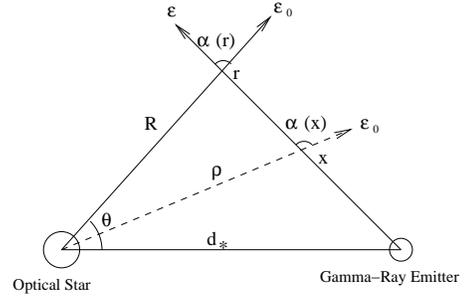}
\caption{Geometry of the photon-photon interaction.}
\label{opac}
\end{center}
\end{figure}

The secondary pair injection function for a monoenergetic point-like source of target photons
is determined by the following
integral:
\begin{eqnarray}\nonumber
 q(\gamma,R)={L_*\over4\pi m_{\rm e}c^3\epsilon_0}\times\quad\quad\quad\quad\quad\quad\quad\quad\quad\quad\quad\quad\quad\quad\quad\quad
\\
\quad\quad\quad\quad\int\limits_{-1}^1{\rm d}
\cos\theta\,{(1-\cos\alpha(r))\over r^2}\int{\rm d} \epsilon\, {{\rm
d}N_\gamma(\epsilon)\over {\rm d}\epsilon{\rm d}t}\,{\rm e}^{-\tau}{{\rm
d}\sigma_{\rm p}\over {\rm d} \gamma}\, ,\quad\quad 
\end{eqnarray}
where ${\rm d}N_\gamma(\epsilon)/{\rm d}\epsilon{\rm d}t$ is the primary
gamma-ray injection spectrum
per time unit, $r$/$\tau$ is the distance/optical depth from the
gamma-ray emitter to the secondary pair creation location, and
$\alpha(r)$ is the interaction angle at this location. A sketch 
of the situation is presented in Fig.~\ref{opac}.
The cross-section  ${\rm
d}\sigma_{\rm p}/ {\rm d} \gamma$ is given be the eq.~(15) from
B\"ottcher \& Schlickeiser~(1997) and the kinematics constraints:
\begin{equation}
{\epsilon\left(1-\sqrt{1-{2\over\epsilon\epsilon_0
(1-\cos\alpha(r))}}\right)\over2}<\gamma<{\epsilon\left(1+\sqrt{1-{2\over\epsilon\epsilon_0
(1-\cos\alpha(r))}}\right)\over2}\,.
\end{equation}
The distance and the interaction angle are defined as the following:
\begin{equation}
r^2=R^2+d_*^2-2d_*R\cos\theta\,,\quad  \cos\alpha(x)={\rho^2+x^2-d_*^2\over2\rho x}\,,
\end{equation}
where $\rho=\sqrt{x^2+d_*^2+x(R^2-d_*^2-r^2)/r}$ is the distance
from the optical star to the gamma-ray absorption point.
%
Finally, the optical depth $\tau$ is calculated like
\begin{equation}
 \tau(r,R)={L_*\over 4\pi m_{\rm e}c^3\epsilon_0}
 \int\limits_0^r{\rm d}x\,
{(1-\cos\alpha(x))\over\rho^2}\sigma_{\rm p}\,,
\end{equation}
where $\sigma_{\rm p}$ is the total pair production cross-section (see e.g.
Berestetskii et al. (\cite{berestetskii82})).

\end{document}